# Stabilizing Single-Atom Catalysts on Metastable Phases of Transition Metal Dichalcogenides

Lina Wang[1,2,3], Zhenhai Wen[1,3*], and Guangfu Luo[4,5,6*]

[1]*State Key Laboratory of Structural Chemistry, and Fujian Provincial Key Laboratory of Materials and Techniques toward Hydrogen Energy, Fujian Institute of Research on the Structure of Matter, Chinese Academy of Sciences, Fuzhou, Fujian 350002, China*

[2]*Fujian Key Laboratory of Green Extraction and High-Value Utilization of New Energy Metals, Fuzhou University, Fuzhou, Fujian 350108, China*

[3]*Fujian Science & Technology Innovation Laboratory for Optoelectronic Information of China, Fuzhou, Fujian 350108, China*

[4]*Department of Materials Science and Engineering, Southern University of Science and Technology, Shenzhen 518055, China*

[5]*Guangdong Provincial Key Laboratory of Computational Science and Material Design, Southern University of Science and Technology, Shenzhen 518055, China*

[6]*Institute of Innovative Materials, Southern University of Science and Technology, Shenzhen 518055, China*

*E-mail: (Z.W.) wen@fjirsm.ac.cn, (G.L.) luogf@sustech.edu.cn

## Abstract

Single-atom catalysts have attracted significant attention due to their exceptional atomic utilization and high efficiency in a range of catalytic reactions. However, these systems often face thermodynamic instability, leading to agglomeration under operational conditions. In this study, we investigate the interactions of twelve types of catalytic atoms (Fe, Co, Ni, Cu, Ru, Rh, Pd, Ag, Ir, Pt, Au, and Bi) on three crystalline phases (1T, 1T′, and 2H) of six transition metal dichalcogenide layers (MoS$_2$, MoSe$_2$, MoTe$_2$, WS$_2$, WSe$_2$, and WTe$_2$) based on first-principles calculations. We ultimately identify 82 stable single-atom systems that thermodynamically prevent the formation of metal clusters on these substrates. Notably, our findings reveal that the metastable 1T and 1T′ phases significantly enhance the binding strength with single atoms and promote their thermodynamic stability. This research offers valuable insights into the design of stable single-atom systems and paves the way for discovering innovative catalysts in the future.

**Keywords**: single-atom catalyst, transition metal dichalcogenide, stability, first-principles calculation

Single-atom catalysts, which consist of isolated atoms dispersed on supporting substrates, have garnered significant attention in recent years due to their outstanding catalytic performance and ultrahigh atomic utilization. These characteristics make them highly suitable for versatile catalytic applications[1,2], such as hydrogen fuel cells[2,3], water electrolysis[4], and carbon dioxide reduction[5,6]. In these catalysts, the supporting substrate plays a crucial role in both anchoring the single atoms and enhancing their catalytic activities[7]. Beyond traditional carbon-based substrates, two-dimensional transitional metal dichalcogenides (TMDs) are emerging as promising platforms for single-atom catalysis[8]. Among TMDs, $MX_2$ materials ($M$ = Mo, W and $X$ = S, Se, Te) are the most widely synthesized using versatile methods, such as hydrothermal grafting, gold hydrogen plasma treatment, and electrochemical techniques[9]. These materials feature diverse chemical compositions, tunable electronic structures, and intrinsic catalytic activity, rendering them ideal candidates as supports for single-atom catalysts[10,11]. Previous applications of these catalysts include electrocatalysis, thermal catalysis, and photocatalysis, with notable examples such as the hydrogen evolution reaction[12,13], hydrogenation reactions[14,15], nitrogen reduction reaction[16], and CO oxidation[17,18].

However, the relatively weak binding between single atoms and substrates often leads to low atomic loading during fabrication and agglomeration under operational conditions, resulting in low active sites and catalyst deactivation[9,19]. Therefore, synthesizing single-atom catalysts with enhanced stability is critical for advancing their industrial applications. Previous studies have identified several approaches to establish strong bonding between metal atoms and their supports[20]. For instance, $CeO_2$ support was employed to anchor isolated Pt atoms due to the relatively strong oxygen-metal bonds[21]. Intrinsic defects in substrates, such as N,S-heteroatom on carbon[22] and Ce vacancy of $CeO_2$[23], have also been leveraged to construct stable single-atom catalysts. Additionally, surface species such as CO were found to protect single atoms from agglomeration or induce particle disintegration[24,25]. Notably, a recent study discovers that different phases of $MoS_2$ can influence single-atom growth. Experiments show that single Pt atoms can be stably dispersed on the 1T′ phase of $MoS_2$[26], showcasing a high loading of 10 wt% and efficient electrocatalytic $H_2$ evolution. In contrast, the 2H phase of $MoS_2$ facilitates the growth of Pt nanoparticles. These findings raise two vital questions: (1) Can these results be generalized to a broader range of single-atom catalysts and TMD substrates? (2) What are the critical factors determining the thermodynamic stability of single-atoms on TMD substrates?

In this study, we investigate the stable configurations of 12 types of single atoms with versatile catalytic applications on the 1T, 1T′, and 2H phases of TMDs ($MX_2$, where $M$ = Mo, W; $X$ = S, Se, Te) using first-principles calculations. Our results reveal 82 novel stable single-atom systems, highlight the crucial role of substrate phase in stabilizing these single atoms, and indicate that certain elements are more favorable for forming single-atom configurations than others.

We examine the thermodynamic stability of 216 single-atom systems, namely 12 types of atoms commonly used in catalysis (Fe, Co, Ni, Cu, Ru, Rh, Pd, Ag, Ir, Pt, Au and Bi) on 18 TMD monolayers (1T, 1T′, and 2H phases of $MoS_2$, $WS_2$, $MoSe_2$, $WSe_2$, $MoTe_2$ and $WTe_2$). As shown in Fig. 1, the 2H phase is characterized by a trigonal prismatic coordination within a hexagonal unit cell and exhibits semiconducting properties. In contrast, the 1T phase features octahedral coordination within a distorted tetragonal unit cell and displays metallic behavior, while the 1T′ phase possesses a distorted octahedral structure and demonstrates semi-metallic characteristics. It is worth noting that the stability of 1T phase is lower than that of 1T′ phase, which in turn is lower than that of the 2H phase (Fig. S1)[27]. For adsorption on the TMD substrates, we coordinate each catalytic atom with one Cl⁻ anion. This is because $TMCl_4^-$ precursor were used in experiments[26,28] and the anions may not be completely removed after adsorption. Besides, both geometry optimization and *ab initio* molecular dynamics simulations indicate that uncoordinated Pt can easily displace S atoms in the 1T′ $MoS_2$ (Figs. S2 and S3), contradicting with the experimental findings of stable Pt single-atoms on the 1T′ $MoS_2$[26], while the Cl⁻ coordinated Pt atoms show high stability.

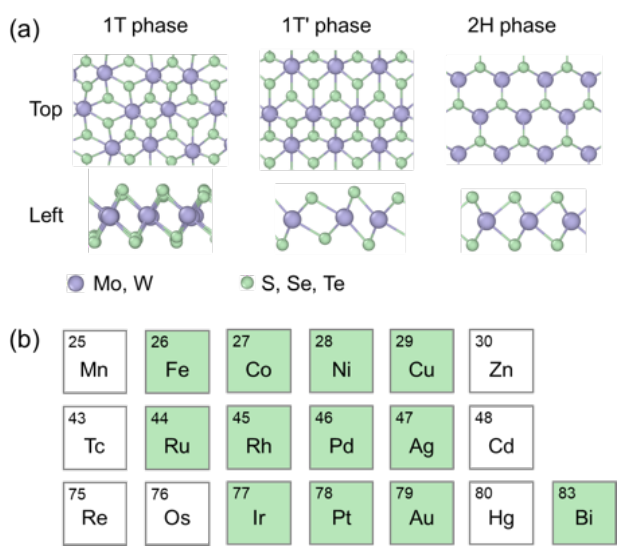

**Figure 1.** (a) Structures of 1T, 1T′, and 2H phase of six TMD material $MX_2$ ($M$ = Mo, W; $X$ = S, Se, Te). (b) Twelve catalytic atoms to be considered for adsorption on $MX_2$.

Our density functional theory (DFT) calculations reveal that the potential adsorption sites for single atoms vary with both substrate and adsorbent. Specifically, the adsorption sites on the 1T phase exhibit four configurations (Fig. 2a): site 1, located above $M$ with bonds to three $X$ atoms; site 2, a bridge site bonded to two $X$ atoms; site 3, positioned above $X$ with one bond to it; and site 4, above $X$ with bonds to three $X$ atoms. The 1T′ phase displays similar adsorption sites as the 1T phase, designated as sites $1^{T'}$, $2^{T'}$, and $3^{T'}$ and $4^{T'}$ (Fig. 2b). For the 2H phase, the corresponding sites are labeled as sites $1^H$, $2^H$, and $3^H$ (Fig. 2c).

To evaluate the stability of single catalytic atoms, we also examine the adsorption of their dimers on the TMD substrates. On the 1T phase, there are four adsorption sites for dimers (Fig. 2d): two neighboring sites formed by site $1^T$ (site $5^T$), two neighboring sites formed by sites $1^T$ and $2^T$ (site $6^T$), two neighboring sites formed by site $3^T$ (site $7^T$), and one site formed by site $3^T$ with the other catalytic atom bonded to Cl$^-$ (site $8^T$). The corresponding sites on the 1T′ phase are designated as sites $5^{T'}$, $6^{T'}$, $7^{T'}$, and $8^{T'}$ (Fig. 2e), while those on the 2H phase are labeled as sites $5^H$, $6^H$, $7^H$, and $8^H$, respectively (Fig. 2f).

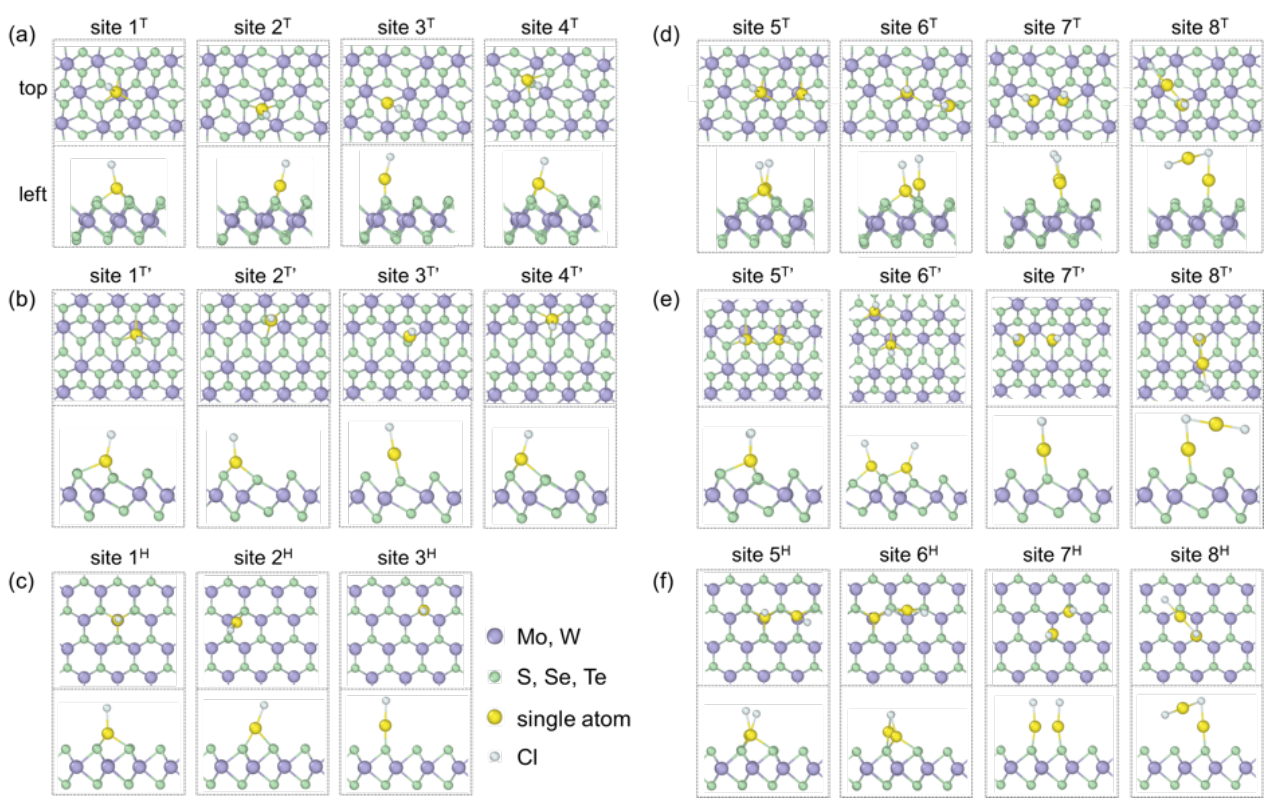

**Figure 2.** Stable or metastable adsorption sites for single-atoms on the (a) 1T, (b) 1T′, and (c) 2H phase of $MX_2$, and those for dimer on the (d) 1T, (e) 1T′, and (f) 2H phase of $MX_2$.

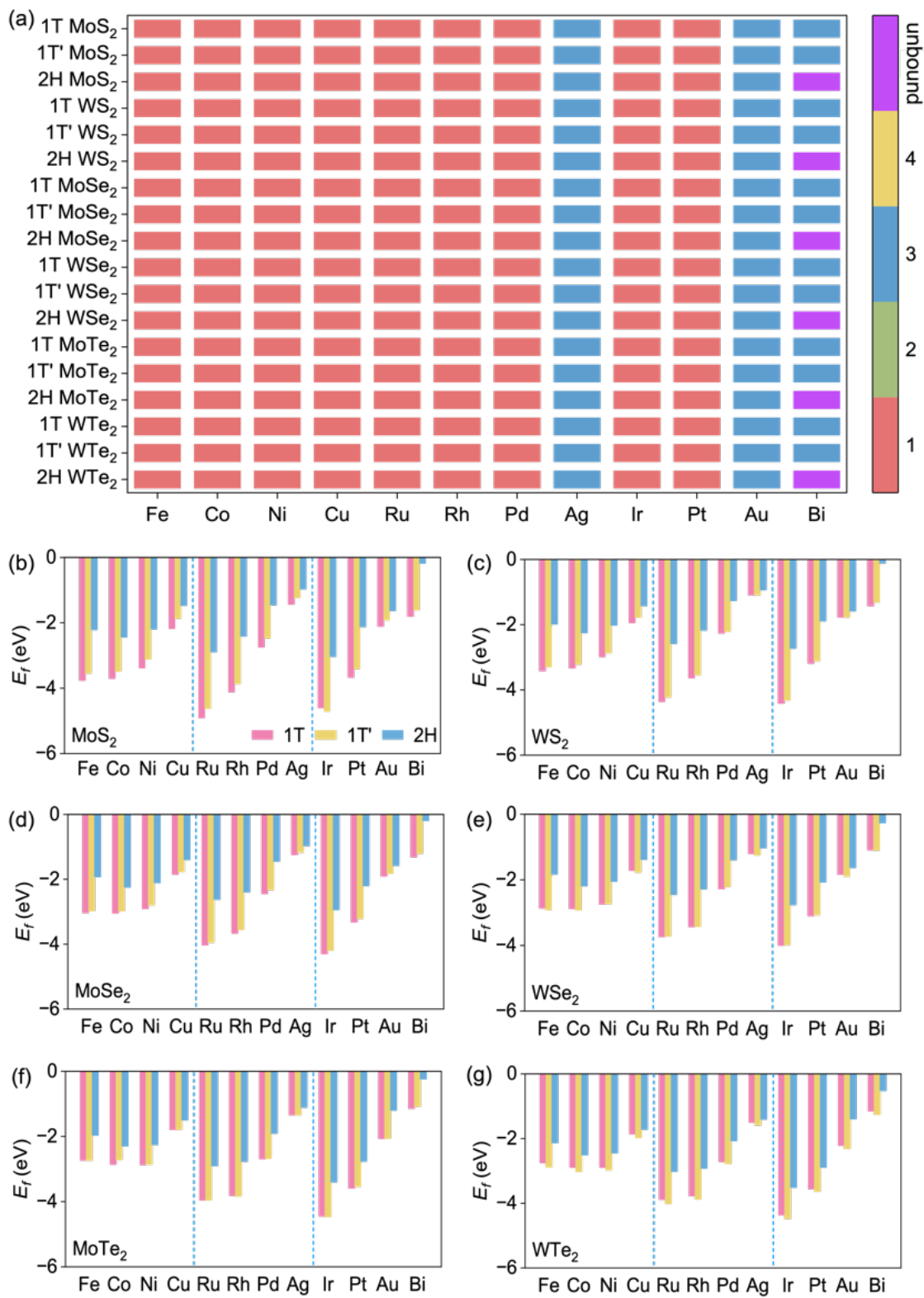

**Figure 3.** (a) Most stable adsorption sites for 12 monomers on $MX_2$ ($M$ = Mo, W; $X$ = S, Se, Te). Different colors represent the absorption sites in Fig. 2a-c. Formation energies of 12 monomers on (b) $MoS_2$, (c) $WS_2$, (d) $MoSe_2$, (e) $WSe_2$, (f) $MoTe_2$, and (g) $WTe_2$.

Figure 3a summarizes the most stable adsorption sites for all 12 catalytic monomers on the 18 types of $MX_2$ (see all structures in Figs. S4-S9 of the Supporting Information). Depending on the adsorbed atoms, the most stable adsorption sites can be categorized into three groups. Specifically, Fe, Co, Ni, Cu, Ru, Rh, Pd, Ir, and Pt monomers favor sites $1^T$, $1^{T'}$, or $1^H$ on the substrates. In contrast, Ag and Au monomers prefer sites $3^T$, $3^{T'}$, or $3^H$ as their most stable adsorption sites. For Bi monomer, it favors sites $3^T$ and $3^{T'}$ on the 1T and 1T′ phases, respectively, while remaining unbound to the 2H phase due to weak adsorption.

The formation energies of monomers exhibit two general features (Fig. 3b-g). In terms of substrate phases, the adsorption strengths on the 1T′ and 1T phases are very close, differing by ~0.05 eV, while adsorption on the 2H phase is significantly (~0.87 eV) weaker than on the two others. This trend aligns with the phase stability order of 1T ~ 1T′ < 2H, indicating that metastable phases are prone to bind more strongly with adsorbents. This phase-dependent binding strength also affects the magnetism of adsorbed monomers. Most monomers on the 1T and 1T' phases show no magnetism, while those on the 2H phases exhibit noticeable magnetism (Figs. S4-S9 of the Supporting Information). This behavior arises because the stronger binding on the 1T and 1T' phases leads to more chemical bonds, which in turn results in fewer unpaired electrons.

Regarding the influences of adsorbed monomers, the fourth-period elements, Fe, Co, and Ni, exhibit similar formation energies, which are notably (~1.02 eV) lower than those of Cu. For monomers in the fifth and sixth periods, the formation energies increase monotonically with atomic number, following the order of Ru, Rh, Pd, and Ag for the fifth period, and Ir, Pr, Au, and Bi for the six period. This trend is generally consistent with the decreasing number of unpaired $d$ electrons, highlighting their crucial roles in the adsorption.

With an understanding of the monomers adsorption, we now explore the stable configurations of catalytic dimers on TMD substrates. As illustrated in Fig. 4a, the Fe, Co, Ni, Ru, Rh, and Ir dimers favor the sites $5^T$, $5^{T'}$, and $5^H$ on the 1T, 1T′, and 2H phases, respectively. The Pd and Pt dimers also exhibit stable adsorption at sites 5T and 5T′ on the 1T and 1T′ phases, respectively. However, the Pd dimer prefers site $6^H$ on the 2H phase, while the Pt dimer favors site $7^H$ on this phase, except 2H MoTe$_2$. In contrast, the Cu, Ag, and Au dimers favor sites $8^T$, $8^{T'}$, and $8^H$ across all the substrates. The Bi dimer remains unbound to all the substrates, with distances to the nearest substrate atoms measuring about 2.9, 3.4, and 3.6 Å for the 1T, 1T′ and 2H phase, respectively. All 216 dimer adsorption structures can be found in Figs. S10-S15 of the Supporting Information.

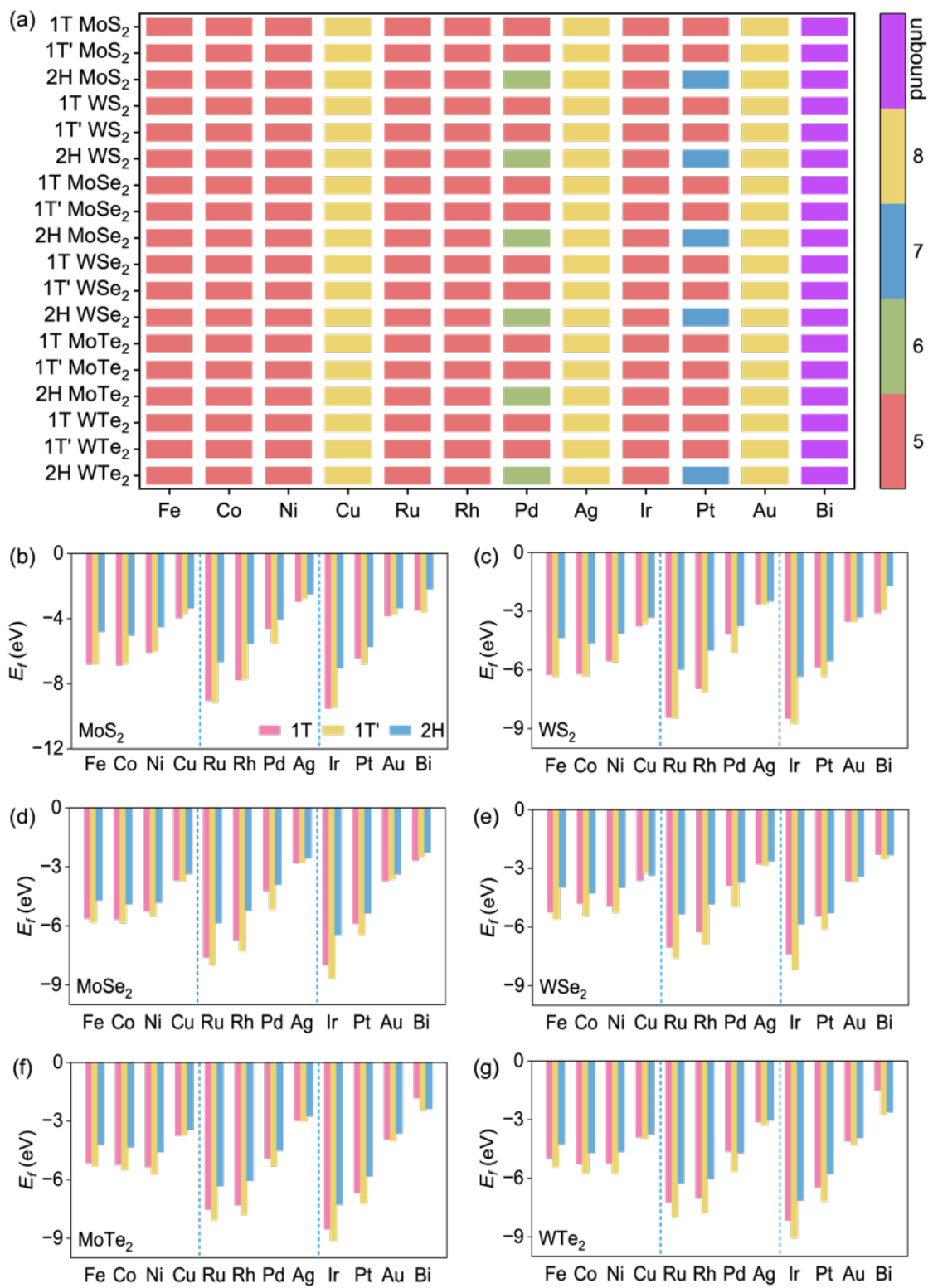

**Figure 4.** (a) Most stable adsorption sites for 12 dimers on $MX_2$ ($M$ = Mo, W; $X$ = S, Se, Te). Different colors represent the absorption sites in Fig. 2d-f. Formation energies of dimers on (b) $MoS_2$, (c) $WS_2$, (d) $MoSe_2$, (e) $WSe_2$, (f) $MoTe_2$, and (g) $WTe_2$.

The formation energy trends of dimers resemble those of monomers (Fig. 4b-g). Specifically, the formation energies of dimers on the 1T and 1T′ phases are comparable, while the 2H phases typically exhibit the weakest adsorption. Among the different dimers, Fe, Co, and Ni exhibit similar formation energies, which are significantly lower than those of Cu. For dimers in the fifth and sixth periods, formation energies increase monotonically in the order of Ru, Rh, Pd, and Ag for the fifth period, and Ir, Pt, Au, and Bi for the six period.

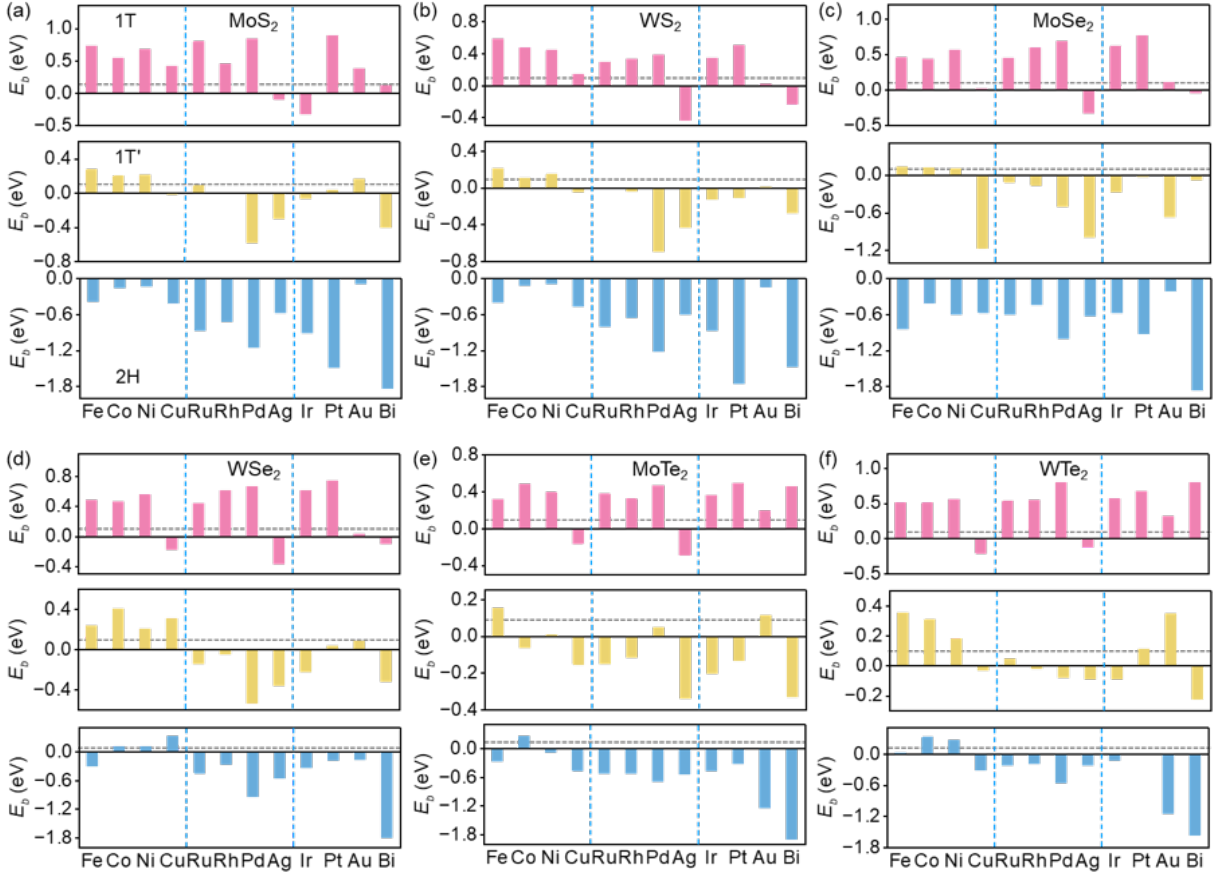

**Figure 5.** Binding energies of dimers relative to monomers on (a) $MoS_2$, (b) $WS_2$, (c) $MoSe_2$, (d) $WSe_2$, (e) $MoTe_2$, and (f) $WTe_2$. The grey horizontal line indicates a binding energy of 0.1 eV.

To evaluate whether the monomers would aggregate to form dimers on the TMD substrates, we examine the dimerization energy, defined as the binding energy of an adsorbed dimer relative to the corresponding adsorbed monomer. As illustrated in Fig. 5, we identify 82 out of the total 216 systems with positive dimerization energies exceeding 0.1 eV, indicating that their dimerization is notably endothermic and thus thermodynamically unfavorable. Moreover, the number of stable monomer systems decreases in the order of 1T, 1T′, and 2H phase, with 56, 20, and 6 systems exhibiting dimerization energy greater than 0.1 eV, respectively. This result suggests that metastable substrates are more likely to stabilize the monomers against dimerization, majorly because their metastable nature leads to a strong adsorption with monomers (Fig. 3b-g). Among the adsorbed atoms, Fe, Co, and Ni display the highest number of stable monomers (each with 12–13 stable systems), followed by Cu, Ru, Rh, Pd, Ir, Pt, and Au, each having 4–7 stable monomer systems. In sharp contrast, Bi has 3 stable monomers, while Ag has none. The instability of Ag single-atom catalyst on all the examined substrates is likely due to its +1 stable charge state, which is saturated by Cl$^-$, thereby hindering its stable binding to the substrate. One potential approach to achieving stable Ag single-atom growth could involve controlling the coordination of anions through the use of different precursors.

In addition to dimerization, we also investigate the binding energies of adsorbed Pt trimer, tetramer, and bulk relative to adsorbed monomer on the 1T, 1T′ and 2H phases of MoS$_2$. Our findings reveal that the endothermic or exothermic nature remains unchanged, while the magnitude of binding energy increases from dimer to trimer and tetramer (Fig. S16 of the Supporting Information). Specifically, the binding energies of Pt dimer, trimer, and tetramer on the 1T MoS$_2$ rise from 0.89 to 1.73 and 2.37 eV, respectively. A similar trend is observed on the 1T′ MoS$_2$. As of the 2H MoS$_2$, the binding energies decrease from -1.48 to -2.46 and -3.52 eV for the dimer, trimer, and tetramer, respectively. The binding energies of Pt bulk relative to the Pt single-atom systems are 0.33, 0.07, and -1.21 eV on the 1T, 1T', and 2H MoS$_2$. Therefore, even with consideration of trimer, tetramer, and even large nanoparticle, Pt monomers are stable on the 1T and 1T' phases of MoS$_2$ but unstable on the 2H phase, aligning well with recent experiments[26]. Additionally, we examine the coordination of NO$_3^-$ with each Pt atom, another potential precursor utilized in experiments[29,30], and find results similar to those observed with Cl$^-$ (Fig. S17 of the Supporting Information). This consistency suggests that the coordination anion may play a minor role in determining the stability of the examined monomers. Since the influences of different phases on stabilizing single atoms is rooted in their thermodynamic behaviors, the key findings of this work are likely applicable to other TMD substrates.

In summary, we investigate the thermodynamic stability of 12 single-atom catalysts (Fe, Co, Ni, Cu, Ru, Rh, Pd, Ag, Ir, Pt, Au, and Bi) supported on 18 TMD substrates (MoS$_2$, WS$_2$, MoSe$_2$, WSe$_2$, MoTe$_2$, and WTe$_2$ in the 1T, 1T′, and 2H phases) through first-principles calculations. We identify 82 systems capable of forming stable monomers that resist dimerization. Notably, the metastable 1T and 1T′ phases are more conducive to the stability of single atoms, while the most stable 2H phase tends to favor the formation of clusters. Among the catalytic atoms, Fe, Co, Ni, Cu, Ru, Rh, Pd, Ir, Pt, and Au are significantly more likely to form stable single atom compared to Bi and Ag. This study reveals new thermodynamically stable single-atom systems and highlights the critical role of substrate phase in stabilizing these configurations.

**Supporting Information**

Computational details; formation energies on all examined substrates; AIMD results of uncoordinated and coordinated Pt monomer on MoS$_2$; most stable adsorbed sites for monomers and dimers on all examined substrates; binding energies of Pt dimer, trimer, and tetramer on MoS$_2$; binding energies of Pt dimer coordinated with Cl$^-$ and NO$_3^-$ anions on MoS$_2$.

**Acknowledgements**

This work was financially supported by the National Foundation of Natural Science, China (No. 52273226), Guangdong Provincial Key Laboratory of Computational Science and Material Design (Grant No. 2019B030301001), the Guangdong Basic and Applied Basic Research Foundation (No. 2024A1515010211), the Shenzhen Science and Technology Innovation Commission (No. JCYJ20200109141412308), the High level of special funds of Southern University of Science and Technology (No. G03050K002), the Postdoctoral Fellowship Program of CPSF, China (GZB20230758), Fujian Key Laboratory of Green Extraction and High-value Utilization of New Energy Metals (No. 2023-KFKT-2), and China Postdoctoral Science Foundation (2024M753237). The calculations were carried out on the Taiyi cluster supported by the Center for Computational Science and Engineering of Southern University of Science and Technology and also on the Major Science and Technology Infrastructure Project of Material Genome Big-science Facilities Platform supported by Municipal Development and Reform Commission of Shenzhen.

**References**

(1) Wang, A. Q.; Li, J.; Zhang, T. Heterogeneous single-atom catalysis. *Nat. Rev. Chem.* **2018**, *2*, 65-81.
(2) Jiang, K.; Back, S.; Akey, A. J.; Xia, C.; Hu, Y. F.; Liang, W. T.; Schaak, D.; Stavitski, E.; Norskov, J. K.; Siahrostami, S.; Wang, H. T. Highly selective oxygen reduction to hydrogen peroxide on transition metal single atom coordination. *Nat. Commun.* **2019**, *10*, 3997.
(3) Li, D. Y.; Zhao, L. L.; Xia, Q.; Wang, J.; Liu, X. M.; Xu, H. R.; Chou, S. L. Activating MoS$_2$ Nanoflakes via Sulfur Defect Engineering Wrapped on CNTs for Stable and Efficient Li-O$_2$ Batteries. *Adv. Funct. Mater.* **2022**, *32*, 2108153.
(4) Wang, X.; Zhang, Y. W.; Si, H. N.; Zhang, Q. H.; Wu, J.; Gao, L.; Wei, X. F.; Sun, Y.; Liao, Q. L.; Zhang, Z.; Ammarah, K.; Gu, L.; Kang, Z.; Zhang, Y. Single-Atom Vacancy Defect to Trigger High-Efficiency Hydrogen Evolution of MoS$_2$. *J. Am. Chem. Soc.* **2020**, *142*, 4298-4308.
(5) Gu, J.; Hsu, C. S.; Bai, L. C.; Chen, H. M.; Hu, X. L. Atomically dispersed Fe sites catalyze efficient CO electroreduction to CO. *Science.* **2019**, *364*, 1091.

(6) Vijay, S.; Ju, W.; Bruckner, S.; Tsang, S. C.; Strasser, P.; Chan, K. R. Unified mechanistic understanding of CO reduction to CO on transition metal and single atom catalysts. *Nat. Catal.* **2021**, *4*, 1024-1031.

(7) Wang, X.; Zhang, Y. W.; Wu, J.; Zhang, Z.; Liao, Q. L.; Kang, Z.; Zhang, Y. Single-Atom Engineering to Ignite 2D Transition Metal Dichalcogenide Based Catalysis: Fundamentals, Progress, and Beyond. *Chem. Rev.* **2022**, *122*, 1273-1348.

(8) Fu, Q.; Han, J. C.; Wang, X. J.; Xu, P.; Yao, T.; Zhong, J.; Zhong, W. W.; Liu, S. W.; Gao, T. L.; Zhang, Z. H.; Xu, L. L.; Song, B. 2D Transition Metal Dichalcogenides: Design, Modulation, and Challenges in Electrocatalysis. *Adv. Mater.* **2021**, *33*, 1907818.

(9) Wang, X.; Zhang, Y.; Wu, J.; Zhang, Z.; Liao, Q.; Kang, Z.; Zhang, Y. Single-Atom Engineering to Ignite 2D Transition Metal Dichalcogenide Based Catalysis: Fundamentals, Progress, and Beyond. *Chem. Rev.* **2022**, *122*, 1273-1348.

(10) Manzeli, S.; Ovchinnikov, D.; Pasquier, D.; Yazyev, O. V.; Kis, A. 2D transition metal dichalcogenides. *Nat. Rev. Mater.* **2017**, *2*, 17033.

(11) Jin, H. Y.; Guo, C. X.; Liu, X.; Liu, J. L.; Vasileff, A.; Jiao, Y.; Zheng, Y.; Qiao, S. Z. Emerging Two-Dimensional Nanomaterials for Electrocatalysis. *Chem. Rev.* **2018**, *118*, 6337-6408.

(12) Zhang, Y. M.; Zhao, J. H.; Wang, H.; Xiao, B.; Zhang, W.; Zhao, X. B.; Lv, T. P.; Thangamuthu, M.; Zhang, J.; Guo, Y.; Ma, J. N.; Lin, L. N.; Tang, J. W.; Huang, R.; Liu, Q. J. Single-atom Cu anchored catalysts for photocatalytic renewable H
production with a quantum efficiency of 56%. *Nat. Commun.* **2022**, *13*.

(13) Lu, Q. P.; Yu, Y. F.; Ma, Q. L.; Chen, B.; Zhang, H. 2D Transition-Metal-Dichalcogenide-Nanosheet-Based Composites for Photocatalytic and Electrocatalytic Hydrogen Evolution Reactions. *Adv. Mater.* **2016**, *28*, 1917-1933.

(14) Li, H. L.; Wang, L. B.; Dai, Y. Z.; Pu, Z. T.; Lao, Z. H.; Chen, Y. W.; Wang, M. L.; Zheng, X. S.; Zhu, J. F.; Zhang, W. H.; Si, R.; Ma, C.; Zeng, J. Synergetic interaction between neighbouring platinum monomers in CO hydrogenation. *Nat. Nanotechnol.* **2018**, *13*, 411-+.

(15) Liu, G. L.; Robertson, A. W.; Li, M. M. J.; Kuo, W. C. H.; Darby, M. T.; Muhieddine, M. H.; Lin, Y. C.; Suenaga, K.; Stamatakis, M.; Warner, J. H.; Tsang, S. C. E. $MoS_2$ monolayer catalyst doped with isolated Co atoms for the hydrodeoxygenation reaction. *Nat. Chem.* **2017**, *9*, 810-816.

(16) Li, J.; Chen, S.; Quan, F. J.; Zhan, G. M.; Jia, F. L.; Ai, Z. H.; Zhang, L. Z. Accelerated Dinitrogen Electroreduction to Ammonia via Interfacial Polarization Triggered by Single-Atom Protrusions. *Chem*. **2020**, *6*, 885-901.

(17) Li, D. L.; Li, W. L.; Zhang, J. P. Al doped $MoS_2$ monolayer: A promising low-cost single atom catalyst for CO oxidation. *Appl Surf Sci*. **2019**, *484*, 1297-1303.

(18) Zhu, Y. D.; Zhao, K.; Shi, J. L.; Ren, X. Y.; Zhao, X. J.; Shang, Y.; Xue, X. L.; Guo, H. Z.; Duan, X. M.; He, H.; Guo, Z. X.; Li, S. F. Strain Engineering of a Defect-Free, Single-Layer $MoS_2$ Substrate for Highly Efficient Single-Atom Catalysis of CO Oxidation. *ACS Appl. Mater. Interfaces*. **2019**, *11*, 32887-32894.

(19) Tang, Y.; Asokan, C.; Xu, M.; Graham, G. W.; Pan, X.; Christopher, P.; Li, J.; Sautet, P. Rh single atoms on TiO(2) dynamically respond to reaction conditions by adapting their site. *Nat. Commun.* **2019**, *10*, 4488.

(20) Tauster, S. J.; Fung, S. C.; Garten, R. L. Strong metal-support interactions. Group 8 noble metals supported on titanium dioxide. *J. Am. Chem. Soc.* **1978**, *100*, 170-175.

(21) Jones, J.; Xiong, H.; DeLaRiva, A. T.; Peterson, E. J.; Pham, H.; Challa, S. R.; Qi, G.; Oh, S.; Wiebenga, M. H.; Hernández, X. I. P.; Wang, Y.; Datye, A. K. Thermally stable single-atom platinum-on-ceria catalysts via atom trapping. *Science*. **2016**, *353*, 150-154.

(22) Zhang, J.; Zhao, Y.; Chen, C.; Huang, Y. C.; Dong, C. L.; Chen, C. J.; Liu, R. S.; Wang, C.; Yan, K.; Li, Y.; Wang, G. Tuning the Coordination Environment in Single-Atom Catalysts to Achieve Highly Efficient Oxygen Reduction Reactions. *J. Am. Chem. Soc.* **2019**, *141*, 20118-20126.

(23) Qiao, B.; Liu, J.; Wang, Y.-G.; Lin, Q.; Liu, X.; Wang, A.; Li, J.; Zhang, T.; Liu, J. Highly Efficient Catalysis of Preferential Oxidation of CO in H2-Rich Stream by Gold Single-Atom Catalysts. *ACS Catalysis*. **2015**, *5*, 6249-6254.

(24) Kauppinen, M. M.; Melander, M. M.; Honkala, K. First-principles insight into CO hindered agglomeration of Rh and Pt single atoms on m-ZrO2. *Catal. Sci. Technol.* **2020**, *10*, 5847-5855.

(25) Liu, J. C.; Wang, Y. G.; Li, J. Toward Rational Design of Oxide-Supported Single-Atom Catalysts: Atomic Dispersion of Gold on Ceria. *J. Am. Chem. Soc.* **2017**, *139*, 6190-6199.

(26) Shi, Z.; Zhang, X.; Lin, X.; Liu, G.; Ling, C.; Xi, S.; Chen, B.; Ge, Y.; Tan, C.; Lai, Z.; Huang, Z.; Ruan, X.; Zhai, L.; Li, L.; Li, Z.; Wang, X.; Nam, G. H.; Liu, J.; He, Q.; Guan, Z.; Wang, J.; Lee, C. S.; Kucernak, A. R. J.; Zhang, H. Phase-dependent growth of Pt on MoS(2) for highly efficient H(2) evolution. *Nature*. **2023**, *621*, 300-305.

(27) Haastrup, S.; Strange, M.; Pandey, M.; Deilmann, T.; Schmidt, P. S.; Hinsche, N. F.; Gjerding, M. N.; Torelli, D.; Larsen, P. M.; Riis-Jensen, A. C.; Gath, J.; Jacobsen, K. W.; Mortensen, J. J.; Olsen, T.; Thygesen, K. S. The Computational 2D Materials Database: high-throughput modeling and discovery of atomically thin crystals. *2d Mater*. **2018**, *5*, 042002.

(28) Gao, Q.; Yao, B. Q.; Pillai, H. S.; Zang, W. J.; Han, X.; Liu, Y. Q.; Yu, S. W.; Yan, Z. H.; Min, B.; Zhang, S.; Zhou, H.; Ma, L.; Xin, H. L.; He, Q.; Zhu, H. Y. Synthesis of core/shell nanocrystals with ordered intermetallic single-atom alloy layers for nitrate electroreduction to ammonia. *Nat. Synth.* **2023**, *2*, 624-634.

(29) Jiao, J. Q.; Lin, R.; Liu, S. J.; Cheong, W. C.; Zhang, C.; Chen, Z.; Pan, Y.; Tang, J. G.; Wu, K. L.; Hung, S. F.; Chen, H. M.; Zheng, L. R.; Lu, Q.; Yang, X.; Xu, B. J.; Xiao, H.; Li, J.; Wang, D. S.; Peng, Q.; Chen, C.; Li, Y. D. Copper atom-pair catalyst anchored on alloy nanowires for selective and efficient electrochemical reduction of CO. *Nat. Chem.* **2019**, *11*, 222-228.

(30) Jiang, K.; Siahrostami, S.; Zheng, T. T.; Hu, Y. F.; Hwang, S.; Stavitski, E.; Peng, Y. D.; Dynes, J.; Gangisetty, M.; Su, D.; Attenkofer, K.; Wang, H. T. Isolated Ni single atoms in graphene nanosheets for high-performance CO reduction. *Energy Environ. Sci.* **2018**, *11*, 893-903.